\documentclass[runningheads]{llncs}

\usepackage{amsmath}
\usepackage{tabularx}
\usepackage[inline,shortlabels]{enumitem}
\usepackage{xcolor}
\setlist[itemize]{noitemsep}
\setlist[enumerate]{noitemsep}
\usepackage[T1]{fontenc}
\usepackage{graphicx}
\usepackage{subcaption}

\usepackage{color}

\usepackage{hyperref}
\hypersetup{
    colorlinks=true,
    linkcolor=blue,
    filecolor=blue,
    urlcolor=blue,
    citecolor=blue
}
\begin{document}

\title{Improving RAG for Personalization with Author Features and Contrastive Examples}

\titlerunning{Personalization with Author Features and Contrastive Examples}

\author{Mert Yazan\inst{1,2}\orcidID{0009-0004-3866-597X} \and
Suzan Verberne \inst{2}\orcidID{0000-0002-9609-9505} \and
Frederik Situmeang \inst{1}\orcidID{0000-0002-2156-2083}}

\authorrunning{Yazan et al.}

\institute{Amsterdam University of Applied Sciences, Fraijlemaborg 133, 1102 CV Amsterdam, Netherlands
\email{\{m.yazan, f.b.i.situmeang\}@hva.nl}\\
\and
Leiden University, Einsteinweg 55, 2333 CC Leiden, Netherlands\\
\email{\{m.yazan, s.verberne\}@liacs.leidenuniv.nl}}

\maketitle              

\begin{abstract}
    Personalization with retrieval-augmented generation (RAG) often fails to capture fine-grained features of authors, making it hard to identify their unique traits. 
    To enrich the RAG context, we propose providing Large Language Models (LLMs) with author-specific features, such as average sentiment polarity and frequently used words, in addition to past samples from the author's profile. We introduce a new feature called \textit{Contrastive Examples}: documents from other authors are retrieved to help LLM identify what makes an author's style unique in comparison to others.   
    Our experiments show that adding a couple of sentences about the named entities, dependency patterns, and words a person uses frequently significantly improves personalized text generation. Combining features with contrastive examples boosts the performance further, achieving a relative 15\% improvement over baseline RAG while outperforming the benchmarks. 
    Our results show the value of fine-grained features for better personalization, while opening a new research dimension for including contrastive examples as a complement with RAG. We release our code publicly.\footnote{ \url{https://github.com/myazann/AP-Bots/tree/main}}
        

\end{abstract}

\section{Introduction}  

    
    Personalized text generation has many application areas, from recommendation \cite{pers_rec} to personalized content creation \cite{ling_control,lamp}. Its implementation with LLMs falls into two groups: via fine-tuning \cite{pieces,democ,hydra}, or via Retrieval-Augmented Generation (RAG) \cite{PEARL,lamp,user_profile}. In fine-tuning-based approaches, the language model is individually adapted for each user \cite{democ}, which remains computationally costly even with parameter-efficient fine-tuning (PEFT) \cite{peft}. RAG-based personalization works by retrieving samples from the author profile, then instructing the LLM to generate personalized content given the samples \cite{lamp}. RAG eliminates the need to fine-tune for each author and outperforms PEFT-based methods when the amount of user data is scarce \cite{lamp_comp,self}.

    The key challenge with RAG is to capture fine-grained author features: retrieving separate, standalone samples from the author's history is not enough to recognize global patterns \cite{pieces}. When presented with previous documents of the author, LLMs tend to memorize certain words and n-grams that are frequently used, instead of understanding the author's style \cite{self}. To address this problem, we derive features from the author's profile. Features enhance the traditional RAG approach by providing global context about the author, nudging the LLM to do more than repeating words and n-grams. We experiment with different types of features, such as the average sentiment polarity, or the words and named entities the person uses the most. In addition, we introduce a new feature called \textit{Contrastive Examples (CE)} which retrieves samples from other authors to better highlight the differences between individual styles. Figure~\ref{fig:framework} gives an overview of our framework. 

    Our results show that adding only a handful of author features alongside samples from the author's profile significantly improves personalization. Furthermore, retrieving contrastive examples helps LLM better understand an author's style. A combination of \textit{CE} and a couple of author features improve the RAG performance by up to \textbf{15\%}. Additionally, our approach does not add any computational overhead and can easily be implemented and deployed. We contribute to the literature by showing that: (a) LLMs can personalize better when presented with author features, (b) contrastive examples helps LLMs distinguish authors, and (c) retrieving contrastive examples can be a tool to differentiate between relevant and irrelevant contexts.

\section{Background and Related Work}

    RAG for personalization is initially implemented by retrieving similar documents from the user's profile, and instructing the LLM to personalize given the documents \cite{lamp}. Later work focuses on different parts of the RAG pipeline, such as optimizing the retriever \cite{PEARL,lamp_optim} or summarizing and synthesizing the documents to improve the reading comprehension of the LLM \cite{teach_personalize}. Alhafni et al. \cite{ling_control} experiment with fine-grained features, such as lexical and syntactic attributes. The crucial difference with our method is that they train smaller models ($\leq$1B) by prepending the attribute vectors to the inputs, instead of RAG. In addition, they do not include features such as the most commonly used words of the author. 

    Even though personalization by its nature is user-specific, finetuning with multiple users before personalizing to one can aid in learning shared features \cite{pieces,hydra}. Tan et al. \cite{pieces} introduce Personalized Pieces, where clusters of users share PEFT parameters, and HYDRA framework proposes an approach to obtain the shared general knowledge between all authors \cite{hydra}. What makes our approach unique is we focus on the differences between authors by retrieving contrastive examples, whereas previous work focused on capturing the similarities. 
    
    The novelty of our idea is not limited to personalized text generation, but can be seen as a new paradigm for retrieval in RAG. Traditionally, RAG is based on retrieving the most relevant documents, and how contrastive examples would work is an open question. A contrastive chain-of-thought approach with both valid and invalid demonstrations performs better than regular chain-of-thought prompts \cite{contrastivecot} and shows that LLMs are capable of making use of contrastive examples. In the context of RAG, retrieving random documents can help with question-answering, while high-scored documents that are not directly relevant hurts the performance \cite{rag_noise}. Since contrastive examples do not fit into either of these categories, their impact on RAG remains unclear. Contrastive learning has previously been studied in Information Retrieval (IR) for training retrievers to differentiate between relevant and irrelevant documents \cite{dpr}. In contrast, our method incorporates both types of documents directly into the LLM's prompt, without modifying the retriever.

\section{Methodology} 

    \begin{figure}[t]
        \centering
        \includegraphics[width=\linewidth,trim={2cm 2cm 0 0},clip]{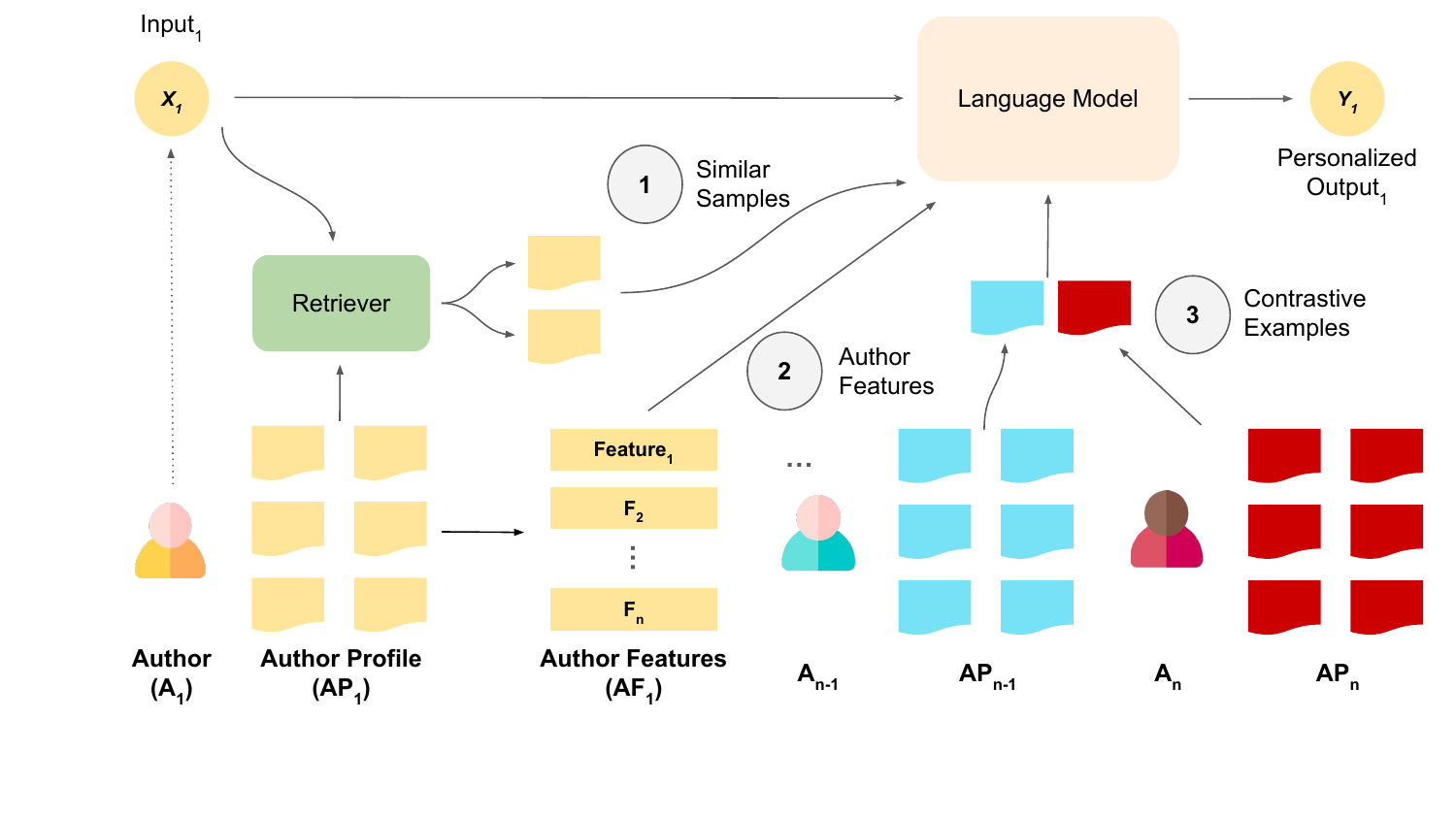}
        \caption{An overview of our approach. The language model receives three inputs: (1) most similar samples from the author profile, given the author's input, (2) features derived from the author profile, (3) contrastive examples gathered from other users. Input (1) denotes the baseline RAG approach, while (2) and (3) are our additions to enrich RAG context.}
        \label{fig:framework}
    \end{figure}

    \paragraph{\textbf{Data}}\label{data} We used 3 datasets offered by the LaMP (When Large Language Models Meet Personalization) benchmark \cite{lamp}: LaMP-4, LaMP-5, and LaMP-7. Since we are focusing on text generation, we excluded the datasets with classification tasks, as well as LaMP-6 because it wasn't accessible. We chose the user-based splits for evaluation. The task in LaMP-4 is to generate a title in the style of the news editor based on a news article given as the input. The author profile consists of previous article-title pairs. The task in LaMP-5 is to generate a title, given the abstract as the input, and the author profile consists of the previous abstract-title pairs of the scholar. Finally, in LaMP-7, the input is a tweet and the task is to paraphrase it given the author's past tweets.

    \paragraph{\textbf{Models}} Given its success in the LaMP benchmark \cite{lamp,self}, we used a pre-trained Contriever \cite{contriever} as the retriever. The average length of the samples differs between datasets, where LaMP-5U has longer documents \cite{lamp}. Therefore, we retrieved 50 documents from author profiles for LaMP-4 and 7, and 7 documents for LaMP-5. Previous research demonstrated that it is not necessary to use that many documents \cite{user_profile,self}, but we see that it is possible to maximize the RAG performance by using more, allowing us to better understand the contribution of our features. 
    
    We started our experiments with OpenChat \cite{OpenChat}, given its success on LaMP \cite{self}. Then we experimented with LLaMA-3.1 \cite{llama3.1} and Gemma-2 \cite{gemma2}. We didn't find drastic variations between models, but Gemma-2 27B performed the best overall. Because of space limitations, we only present its results in Section \ref{results}. We didn't change the default generation parameters on HuggingFace, except for the \textit{temperature} and \textit{max\_new\_tokens} (maximum number of generation tokens). Given the shortness of the expected generations, we set \textit{max\_new\_tokens} to 128 tokens, and \textit{temperature} is set to 0.7. We noticed negligible changes in the final result with different generation parameters, so we decided to use the default ones. 

    We constructed the prompts by first giving a role to the LLM (\textit{you are a news editor}), and explaining the types of inputs it will receive. We put similar documents from the author's past as the first input. Then we include features and contrastive examples (see Figure~\ref{fig:framework}). Similar past documents are always included as input, since they form the baseline RAG. Finally, we instruct the model to generate the proper text (e.g. title) given the features it received. The prompt structure is consistent for each dataset, deviating only on the role and the instruction. We didn't see an improvement using Chain-of-Thought \cite{cot} or similar variants, while noticing an exponential increase in inference times. Therefore, we didn't use such prompting methods. Prompts are available in our public repository.
        
    \paragraph{\textbf{Features}} Our features can be categorized into two: (1)\textit{ Author Features}, derived from the author's profile and (2) \textit{Contrastive Examples}, which are samples coming from other authors. 

    \paragraph{Author Features:} 
    
    We define the following features: 
    
    \begin{itemize}
          \item \textbf{SP:} Average sentiment polarity (\(SP \in [-1, 1]\))
          \item \textbf{SUBJ:} Average subjectivity (\(SUBJ \in [0, 1]\))
          \item \textbf{SMOG:} SMOG Index\footnote{SMOG is a measure of readability and typically has a value between 1-12, estimating the years of education needed to understand a piece of writing.}
          \item \textbf{ADVU:} Adverb usage percentage on average (\(ADVU \in [0, 100]\))
          \item \textbf{ADJU:} Adjective usage percentage on average (\(ADJU \in [0, 100]\))
          \item \textbf{PU:} Pronoun usage percentage on average (\(PU \in [0, 100]\))
          \item \textbf{NEF:} Most frequently used named entities (\([ne_0, ..., ne_9]\))
          \item \textbf{DPF:} Most frequently used dependency patterns (\([dp_0, ..., dp_9]\))
          \item \textbf{WF:} Most frequently used words (\([w_0, ..., w_9]\))

    \end{itemize}
    
    \noindent    We incorporated the features into the prompt by converting their definition into the following sentence format: \textit{\{feat\_def\} for the writer is \{feat\_value\}}. We use TextBlob \cite{textblob} to obtain the subjectivity and sentiment polarity scores, and textstat \cite{textstat} for the SMOG index. We use NLTK \cite{nltk} for calculating \textit{ADVU, ADJU, PU} and \textit{WF}. Finally, we use spaCy \cite{spacy2} to get named entities and dependency patterns. The averages are calculated by taking the mean of the feature values for all the samples in the author profile. We choose the top 10 most frequent elements for frequency features, since our experiments showed that more than 10 does not change the performance. 
    
    \paragraph{Contrastive Examples (CE):}

    Contrastive examples can be considered hard negatives, as we choose authors that are the least similar to the current author. We calculate the distances between authors using the inputs (see the Data paragraph for the inputs of each dataset) with Contriever. Alternatively, we tried looking at the similarity between author profiles based on the average embeddings. We didn't find a significant difference, therefore we proceeded with using the inputs. After the least similar authors are identified, samples from their profiles are chosen randomly. The number of samples coming from other authors were fixed: 1 sample per each author for LaMP-5, and 3 samples for LaMP-4 and LaMP-7. We use fewer samples compared to how many we retrieve from the current author's profile because of context window limitations and saturated performance with fewer samples. In LaMP-5, using too many contrastive examples even hurts the performance. We suspect the length of the samples in that dataset causes noise and decreases the LLMs' attention to the different parts of the context.

     \begin{figure}[t]
        \centering
        \includegraphics[width=\linewidth]{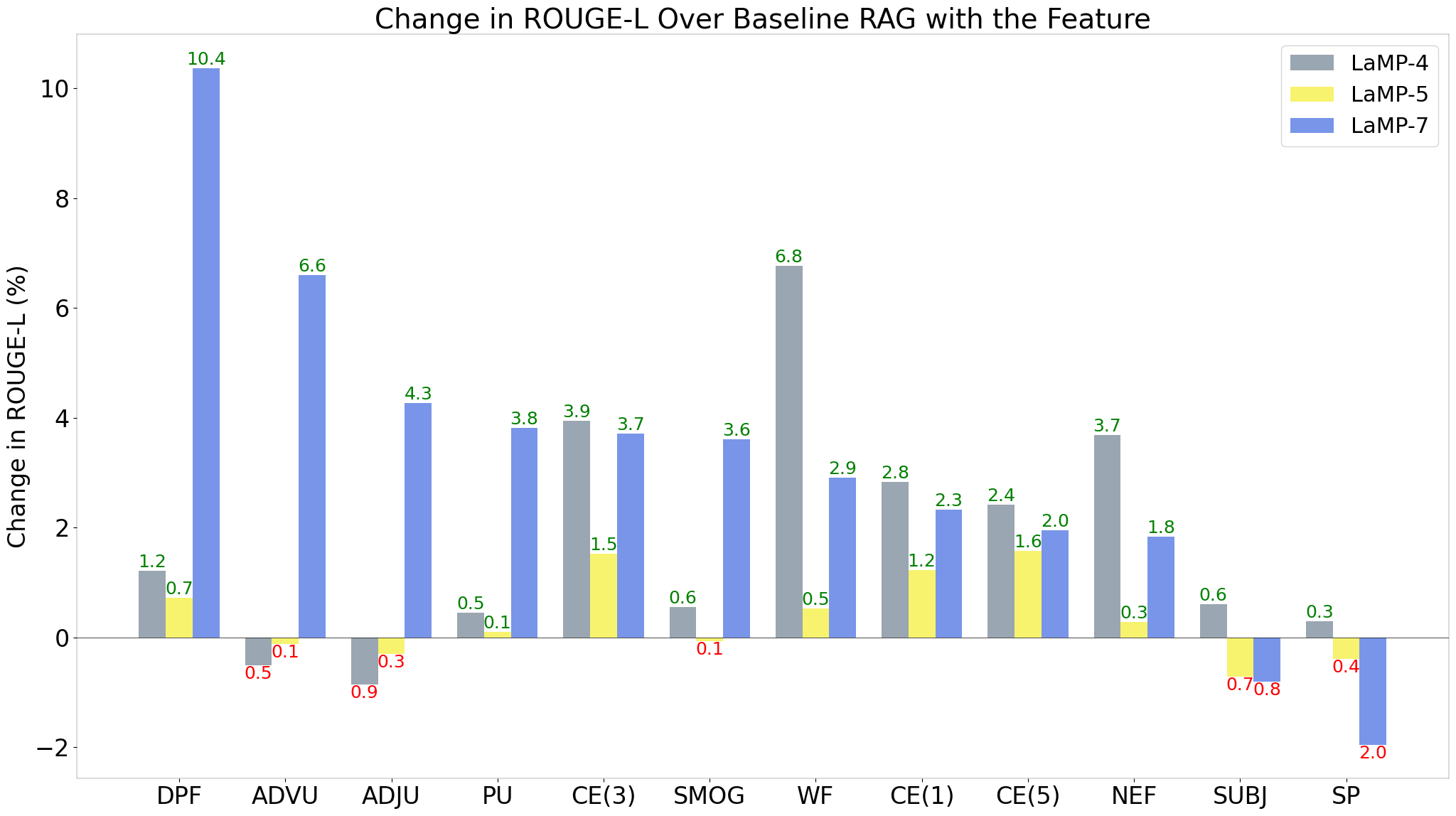}
        \caption{The improvement each feature provides for Rouge-L on validation sets when included on top of baseline RAG. In this case, features are not combined but used individually. A negative change signifies the feature hurting the baseline. Values inside parentheses show the number of contrastive users.}
        \label{fig:single_features}
    \end{figure}

\section{Results}\label{results}

    Our baseline is a RAG experiment without any author features. Then, we test each feature individually on validation sets to understand how much they contribute to the baseline. Figure~\ref{fig:single_features} shows that some features improve the baseline for every dataset, while others hurt. \textit{CE} emerges as a robust feature since it offers at least a 1\% increase regardless of the dataset. Retrieving samples from 5 different authors performs worse than 3, possibly due to increased noise. LaMP-7 benefits the most from the features, \textit{DPF} itself can increase Rouge-L by more than \textbf{10\%}. Since LaMP-7 is about tweets, which are highly personal and informal, dependency patterns better highlight individual's styles compared to others. Compared to tweets, news articles are more formal and contain domain-specific keywords. Therefore, \textit{WF} becomes a more suitable feature for LaMP-4 instead of \textit{DPF}. Performance gains in LaMP-5 are more limited. Authors in LaMP-5 represent scholars, each working in a very distinct and highly specific academic domain. Also, documents in LaMP-5 are much  longer than other datasets. Therefore, retrieved samples contain more information compared to other datasets, making it hard for author features to contribute. In comparison, contrastive examples highlight the vast differences between specialized academic domains, becoming the best performing feature in LaMP-5.

    Overall, features presented as numerical values (\textit{SP, SUBJ, ADVU, ADJU, PU}) are less effective or even hurtful compared to ones presented as words (\textit{CE, WF, NEF, DPF}). This may be due to the LLMs shortcoming of working with numerical values \cite{numerologic}. In our qualitative analysis, we noticed LLMs overly polarizing (e.g. too negative or too positive) their outputs in the presence of numerical features. They especially have a hard time adjusting to subjectivity and sentiment polarity, not being able to capture the nuances of scores that are closer to the mean. In fairness, it is hard to conceptualize the difference between a sentiment score of 0.29 and 0.4, especially given the personal differences in expression.
    
    \begin{table}[t]       
        \begin{tabular}{l|c|c|c|c|c|c}
        \multicolumn{1}{c}{} & \multicolumn{2}{c|}{\textbf{LaMP-4}} & \multicolumn{2}{c|}{\textbf{LaMP-5}} & \multicolumn{2}{c}{\textbf{LaMP-7}} \\
         & Rouge-1 & Rouge-L & Rouge-1 & Rouge-L & Rouge-1 & Rouge-L \\
        \hline
        Finetuned FlanT5-base \cite{lamp} & 0.186 & 0.171 & 0.450 & 0.409 & \textbf{0.528} & \textbf{0.475} \\
        \hline
        $\text{Learning to Prompt}^*$ \cite{learningtoprompt} & 0.221 & 0.202 & 0.472 & 0.432 & - & - \\
        \hline
        RAG & 0.214 & 0.196 & 0.475 & 0.425 & 0.437 & 0.380 \\
        \hline
        RAG  + (WF, CE(3)) & \textbf{0.228} & \textbf{0.210} & 0.481 & 0.431 & 0.483 & 0.431 \\
        \hline
        RAG + (DPF, CE(3)) & 0.226 & 0.207 & \textbf{0.483} & \textbf{0.432} & 0.496 & 0.437 \\ 
        \hline
        RAG + (WF, DPF, CE(3)) & 0.221 & 0.202 & 0.481 & 0.430 & 0.501 & 0.441 \\
        \hline
        \end{tabular}
        \caption{Our best feature combinations compared to the current leaderboard for user-based test splits. 
        $^*$ The method \cite{learningtoprompt} is based on finetuning a FlanT5-XXL with PEFT.}
        \label{tab:benchmark_comp}
    \end{table}
    
    Next, we evaluate the effect of combining features. We chose the best combinations from the validation set results, and Table~\ref{tab:benchmark_comp} shows their performance on test sets (for brevity, we present only the combinations that work best for each dataset). Combining features significantly\footnote{Paired t-tests between baseline RAG and the best performing methods with $\alpha=0.05$ confirmed the statistical significance: LaMP-4: ($p = 2.7 \times 10^{-11}$), LaMP-5: ($p = 0.0014$), LaMP-7: ($p = 9.6\times 10^{-32}$). Since test set predictions are not accessible, t-tests are conducted on the validation sets.} improves the performance. \textit{CE} emerges as a great complementary feature, as it is present in all the best performing combinations. Compared with the leaderboard\footnote{\url{https://lamp-benchmark.github.io/leaderboard}}, our approach surpasses the top methods in LaMP-4 and LaMP-5. Since the other methods are finetuning-based, we can say that RAG-based personalization combined with our features can eliminate the need to finetune for each author. Contrarily, finetuning comes out on top in LaMP-7. Still, our method reduces the gap between fine-tuning and RAG by boosting the baseline RAG performance by \textbf{15\%} relatively (from 0.437 to 0.501). Since LaMP-7 benefitted the most from the features, we believe there is still room for improvement to improve the performance with more features.
    
\section{Conclusion}

    Our research pushes the boundaries of personalization by enhancing the retrieval component of RAG with fine-grained author features and contrastive examples. We show that a single sentence providing which words an author utilizes the most combined with contrastive examples can improve the RAG performance by \textbf{15\%}. Our research offers a new paradigm for RAG, where contrastive examples complement the retrieved samples and help LLM better identify the ground truth. For future work, more suitable ways to present numerical features can be explored. More advanced methods to mitigate the limitations of presenting numerical features to LLMs might be a promising direction for exploration. How to retrieve contrastive examples is an open question for the IR community, as well as extending the contrastive retrieval approach with other IR tasks.

\begin{credits}


\end{credits}



\bibliography{mainbib}

\begin{thebibliography}{10}
\providecommand{\url}[1]{\texttt{#1}}
\providecommand{\urlprefix}{URL }
\providecommand{\doi}[1]{https://doi.org/#1}

\bibitem{ling_control}
Alhafni, B., Kulkarni, V., Kumar, D., Raheja, V.: Personalized text generation with fine-grained linguistic control. In: Deshpande, A., Hwang, E., Murahari, V., Park, J.S., Yang, D., Sabharwal, A., Narasimhan, K., Kalyan, A. (eds.) Proceedings of the 1st Workshop on Personalization of Generative AI Systems (PERSONALIZE 2024). pp. 88--101. Association for Computational Linguistics, St. Julians, Malta (Mar 2024), \url{https://aclanthology.org/2024.personalize-1.8}

\bibitem{nltk}
Bird, S., Klein, E., Loper, E.: Natural language processing with Python: analyzing text with the natural language toolkit. " O'Reilly Media, Inc." (2009)

\bibitem{contrastivecot}
Chia, Y.K., Chen, G., Luu, A.T., Poria, S., Bing, L.: Contrastive chain-of-thought prompting. ArXiv  \textbf{abs/2311.09277} (2023), \url{https://api.semanticscholar.org/CorpusID:265221368}

\bibitem{rag_noise}
Cuconasu, F., Trappolini, G., Siciliano, F., Filice, S., Campagnano, C., Maarek, Y., Tonellotto, N., Silvestri, F.: The power of noise: Redefining retrieval for rag systems. In: Proceedings of the 47th International ACM SIGIR Conference on Research and Development in Information Retrieval. p. 719–729. SIGIR '24, Association for Computing Machinery, New York, NY, USA (2024). \doi{10.1145/3626772.3657834}, \url{https://doi.org/10.1145/3626772.3657834}

\bibitem{llama3.1}
Dubey, A., Jauhri, A., Pandey, A., et~al.: The llama 3 herd of models (2024), \url{https://arxiv.org/abs/2407.21783}

\bibitem{spacy2}
Honnibal, M., Montani, I.: {spaCy 2}: Natural language understanding with {B}loom embeddings, convolutional neural networks and incremental parsing (2017), to appear

\bibitem{peft}
Houlsby, N., Giurgiu, A., Jastrzebski, S., Morrone, B., de~Laroussilhe, Q., Gesmundo, A., Attariyan, M., Gelly, S.: Parameter-efficient transfer learning for nlp. ArXiv  \textbf{abs/1902.00751} (2019), \url{https://api.semanticscholar.org/CorpusID:59599816}

\bibitem{contriever}
Izacard, G., Caron, M., Hosseini, L., Riedel, S., Bojanowski, P., Joulin, A., Grave, E.: Unsupervised dense information retrieval with contrastive learning (2022), \url{https://arxiv.org/abs/2112.09118}

\bibitem{dpr}
Karpukhin, V., Oguz, B., Min, S., Lewis, P., Wu, L., Edunov, S., Chen, D., Yih, W.t.: Dense passage retrieval for open-domain question answering. In: Webber, B., Cohn, T., He, Y., Liu, Y. (eds.) Proceedings of the 2020 Conference on Empirical Methods in Natural Language Processing (EMNLP). pp. 6769--6781. Association for Computational Linguistics, Online (Nov 2020). \doi{10.18653/v1/2020.emnlp-main.550}, \url{https://aclanthology.org/2020.emnlp-main.550}

\bibitem{teach_personalize}
Li, C., Zhang, M., Mei, Q., Wang, Y., Hombaiah, S.A., Liang, Y., Bendersky, M.: Teach llms to personalize -- an approach inspired by writing education (2023), \url{https://arxiv.org/abs/2308.07968}

\bibitem{textblob}
Loria, S.: \url{https://textblob.readthedocs.io}

\bibitem{learningtoprompt}
Mineiro, P.: \url{https://github.com/pmineiro/lampstuff}

\bibitem{PEARL}
Mysore, S., Lu, Z., Wan, M., Yang, L., Menezes, S., Baghaee, T., Gonzalez, E.B., Neville, J., Safavi, T.: Pearl: Personalizing large language model writing assistants with generation-calibrated retrievers. ArXiv  \textbf{abs/2311.09180} (2023), \url{https://api.semanticscholar.org/CorpusID:265213422}

\bibitem{lamp_optim}
Salemi, A., Kallumadi, S., Zamani, H.: Optimization methods for personalizing large language models through retrieval augmentation. In: Proceedings of the 47th International ACM SIGIR Conference on Research and Development in Information Retrieval. p. 752–762. SIGIR '24, Association for Computing Machinery, New York, NY, USA (2024). \doi{10.1145/3626772.3657783}, \url{https://doi.org/10.1145/3626772.3657783}

\bibitem{lamp}
Salemi, A., Mysore, S., Bendersky, M., Zamani, H.: {L}a{MP}: When large language models meet personalization. In: Ku, L.W., Martins, A., Srikumar, V. (eds.) Proceedings of the 62nd Annual Meeting of the Association for Computational Linguistics (Volume 1: Long Papers). pp. 7370--7392. Association for Computational Linguistics, Bangkok, Thailand (Aug 2024). \doi{10.18653/v1/2024.acl-long.399}, \url{https://aclanthology.org/2024.acl-long.399}

\bibitem{lamp_comp}
Salemi, A., Zamani, H.: Comparing retrieval-augmentation and parameter-efficient fine-tuning for privacy-preserving personalization of large language models. arXiv preprint arXiv:2409.09510  (2024)

\bibitem{numerologic}
Schwartz, E., Choshen, L., Shtok, J., Doveh, S., Karlinsky, L., Arbelle, A.: Numerologic: Number encoding for enhanced llms' numerical reasoning (2024), \url{https://arxiv.org/abs/2404.00459}

\bibitem{pieces}
Tan, Z., Liu, Z., Jiang, M.: Personalized pieces: Efficient personalized large language models through collaborative efforts (2024), \url{https://arxiv.org/abs/2406.10471}

\bibitem{democ}
Tan, Z., Zeng, Q., Tian, Y., Liu, Z., Yin, B., Jiang, M.: Democratizing large language models via personalized parameter-efficient fine-tuning (2024), \url{https://arxiv.org/abs/2402.04401}

\bibitem{gemma2}
Team, G., Riviere, M., Pathak, S., et~al.: Gemma 2: Improving open language models at a practical size (2024), \url{https://arxiv.org/abs/2408.00118}

\bibitem{textstat}
Textstat: \url{https://textstat.org}

\bibitem{OpenChat}
Wang, G., Cheng, S., Zhan, X., Li, X., Song, S., Liu, Y.: Openchat: Advancing open-source language models with mixed-quality data. ArXiv  \textbf{abs/2309.11235} (2023), \url{https://api.semanticscholar.org/CorpusID:262064307}

\bibitem{cot}
Wei, J., Wang, X., Schuurmans, D., Bosma, M., Ichter, B., Xia, F., Chi, E.H., Le, Q.V., Zhou, D.: Chain-of-thought prompting elicits reasoning in large language models. In: Proceedings of the 36th International Conference on Neural Information Processing Systems. NIPS '22, Curran Associates Inc., Red Hook, NY, USA (2024)

\bibitem{user_profile}
Wu, B., Shi, Z., Rahmani, H.A., Ramineni, V., Yilmaz, E.: Understanding the role of user profile in the personalization of large language models (2024), \url{https://arxiv.org/abs/2406.17803}

\bibitem{pers_rec}
Wu, L., Zheng, Z., Qiu, Z., Wang, H., Gu, H., Shen, T., Qin, C., Zhu, C., Zhu, H., Liu, Q., Xiong, H., Chen, E.: A survey on large language models for recommendation. World Wide Web  \textbf{27}(5) (Aug 2024). \doi{10.1007/s11280-024-01291-2}, \url{https://doi.org/10.1007/s11280-024-01291-2}

\bibitem{self}
Yazan, M., Verberne, S., Situmeang, F.: The impact of quantization on retrieval-augmented generation: An analysis of small llms (2024), \url{https://arxiv.org/abs/2406.10251}

\bibitem{hydra}
Zhuang, Y., Sun, H., Yu, Y., Qiang, R., Wang, Q., Zhang, C., Dai, B.: Hydra: Model factorization framework for black-box llm personalization (2024), \url{https://arxiv.org/abs/2406.02888}

\end{thebibliography}
\bibliographystyle{splncs04}

\end{document}